\documentclass[11pt]{article}
\usepackage{amsmath,amssymb,graphicx}
\usepackage{hyperref}

\usepackage{cite}
\usepackage{color}
\usepackage{dsfont}

\definecolor{darkred}{rgb}{0.65,0.15,0}
\hypersetup{pdfborder={0 0 0},colorlinks=true,urlcolor=darkred,citecolor=blue,linkcolor=darkred,linktocpage=true}

\newcommand{\doi}[2]{\href{http://dx.doi.org/#2}{#1}}

\newcommand{\nn}{\nonumber}

\newcommand{\ints}{\mathbb{Z}}
\newcommand{\reals}{\mathbb{R}}
\newcommand{\cx}{\mathbb{C}}

\newcommand{\ME}{\mathbb{E}}

\newcommand{\UHP}{\mathbb{H}}
\newcommand{\gs}{g_{\mathrm{s}}}

\newcommand{\mf}[1]{{\mathfrak{#1}}}

\newcommand{\I}{\mathds{1}}

\makeatletter

\@addtoreset{equation}{section}
\makeatother

\begin{document}

\thispagestyle{empty}

\mbox{ }
\vspace{30mm}

\begin{center}
{\LARGE \bf $D^6R^4$ curvature corrections, modular graph functions\\[5mm] and Poincar\'e series}\\[10mm]

\vspace{8mm}
\normalsize
{\large  Olof Ahl\'en${}^{1}$ and Axel Kleinschmidt${}^{1,2}$}

\vspace{10mm}
${}^1${\it Max-Planck-Institut f\"{u}r Gravitationsphysik (Albert-Einstein-Institut)\\
Am M\"{u}hlenberg 1, DE-14476 Potsdam, Germany}
\vskip 1 em
${}^2${\it International Solvay Institutes\\
ULB-Campus Plaine CP231, BE-1050 Brussels, Belgium}

\vspace{20mm}

\hrule

\vspace{10mm}

\begin{tabular}{p{12cm}}
{\small
In this note we study the U-duality invariant coefficient functions of higher curvature corrections to the four-graviton scattering amplitude in type IIB string theory compactified on a torus. The main focus is on the $D^6R^4$ term that is known to satisfy an inhomogeneous Laplace equation. We exhibit a novel method for solving this equation in terms of a Poincar\'e series ansatz and recover known results in $D=10$ dimensions and find new results in $D<10$ dimensions.  We also apply the method to modular graph functions as they arise from closed superstring one-loop amplitudes.
}
\end{tabular}

\vspace{10mm}
\hrule

\end{center}

\newpage

\setcounter{page}{1}

\tableofcontents

\section{Introduction}

String theory predicts very precise corrections to field theory scattering amplitudes via the $\alpha'$-expansion~\cite{Gross:1986iv}, where $\alpha'=\ell_{\mathrm{s}}^2$ is related to the inverse string tension. Via this process the contribution of massive string states to the scattering of massless particles can be systematically evaluated. Using further the constraints implied by U-duality and supersymmetry one can sometimes even determine exactly the perturbative and non-perturbative contributions to the string scattering amplitude at a fixed order in $\alpha'$. This strategy was first employed by Green and Gutperle~\cite{Green:1997tv} and further developed in many subsequent papers, see for example~\cite{Green:1997as,Kiritsis:1997em,Pioline:1998mn,Green:1998by,Obers:1999um,Green:1999pu,Pioline:2001jn,Green:2005ba,Basu:2007ru,Basu:2007ck,Green:2008uj,Green:2010wi,Pioline:2010kb,Gubay:2010nd,Green:2010kv,Basu:2011he,Green:2011vz,Gubay:2012sy,Fleig:2012xa,Garousi:2013tca,Bossard:2014lra,Bossard:2014aea,Pioline:2015yea,Wang:2015jna,Bossard:2015uga,Wang:2015aua,Bossard:2015foa,Bossard:2017kfv}.

The structure of the (analytic contribution to the) four-graviton amplitude can be written in terms of an effective low-energy theory in $D$ space-time dimensions (in Einstein frame) as
\begin{align}
\mathcal{L}^{(D)} \sim \frac{1}{(\alpha')^3}R + \sum_{p,q=0}^\infty (\alpha')^{2p+3q} \mathcal{E}_{(p,q)}^{(D)}(g) D^{4p+6q} R^4 + \ldots,
\end{align}
where we have suppressed an overall dimensionful factor. There is a double summation over two integers due to the two independent symmetric invariants\footnote{The dimensionless Mandelstam invariants are here defined as $s=-\frac{\alpha'}{4}(k_1+k_2)^2$, $t=-\frac{\alpha'}{4}(k_1+k_4)^2$ and $u=-\frac{\alpha'}{4}(k_1+k_3)^2$ and satisfy $s+t+u=0$ on-shell.}
\begin{align}
\sigma_2 = s^2+t^2+u^2, \quad \sigma_3 = s^3+t^3+u^3
\end{align}
that can be constructed for the scattering of four massless particles and that can appear in the scattering amplitude. The term $D^{4p+6q} R^4$ is short-hand for very specific contractions of $(4p+6q)$ covariant derivatives of four Riemann tensors. The main interest lies in the functions $\mathcal{E}_{(p,q)}^{(D)}(g)$ of the moduli $g\in E_{11-D}(\reals)/K(E_{11-D})$, where $E_{11-D}$ is the (split real) Cremmer--Julia hidden symmetry group of maximal ungauged supergravity in $D$ space-time dimensions~\cite{Cremmer:1979up} and $K(E_{11-D})$ its maximal compact subgroup. (The Dynkin diagram of $E_{11-D}$ is given below in figure~\ref{fig:dynk}.)

The functions have to be invariant under the discrete U-duality $E_{11-D}(\ints)$~\cite{Hull:1994ys} and supersymmetry implies that they also have to satisfy (tensorial) differential equations~\cite{Green:1998by,Green:2005ba,Bossard:2014lra,Bossard:2015uga}. The most prominent of these are the second order Poisson-type equations for the first three terms in the expansion~\cite{Green:1998by,Green:2005ba,Pioline:2015yea}
\begin{subequations}
\label{eq:Lapl}
\begin{align}
\left(\Delta - \frac{3(11-D)(D-8)}{D-2} \right) \mathcal{E}^{(D)}_{(0,0)}&= 6\pi\delta_{D,8},&\\
\left(\Delta - \frac{5(12-D)(D-7)}{D-2} \right) \mathcal{E}^{(D)}_{(1,0)} &= 40\zeta(2)\delta_{D,7} + 7\mathcal{E}_{(0,0)}^{(4)} \delta_{D,4},&\\
\label{eq:D6R4}
\left(\Delta - \frac{6(14-D)(D-6)}{D-2} \right) \mathcal{E}^{(D)}_{(0,1)} &= -\left(\mathcal{E}_{(0,0)}^{(D)}\right)^2+40\zeta(3)\delta_{D,6}+\frac{55}{3}\mathcal{E}_{(0,0)}^{(5)} \delta_{D,5}+\frac{85}{2\pi}\mathcal{E}_{(1,0)}^{(4)}\delta_{D,4},&
\end{align}
\end{subequations}
where $\Delta$ is the Laplace--Beltrami operator on the classical moduli space $E_{11-D}/K(E_{11-D})$. The Kronecker delta terms on the right-hand sides of these equations are related to divergences in supergravity in specific dimensions whose treatment is known~\cite{Green:2010wi,Green:2010sp,Bossard:2015oxa}. Disregarding these, the first two equations in~\eqref{eq:Lapl} are homogeneous Laplace equations for the two functions $\mathcal{E}_{(0,0)}^{(D)}$ and $\mathcal{E}_{(1,0)}$ that correspond to the $\tfrac12$-BPS $R^4$ and $\tfrac14$-BPS $D^4R^4$ curvature corrections, respectively. Their solutions have been studied in great detail and are given by (linear combinations of) Eisenstein series, see for instance~\cite{Green:2010wi,Pioline:2010kb,Green:2010kv,Pioline:2015yea,Fleig:2015vky} for summaries or section~\ref{sec:R4D4R4} below. By contrast, equation~\eqref{eq:D6R4} for $\mathcal{E}_{(0,1)}^{(D)}$ (that corresponds to the $\tfrac18$-BPS correction $D^6R^4$) is of a qualitatively different nature in that it always contains a non-linear source term given by $-(\mathcal{E}_{(0,0)}^{(D)})^2$. The equation is therefore an inhomogeneous Laplace equation or Poisson equation. A similar structure is expected for curvature corrections with even more derivatives~\cite{Green:2008bf}. We also note that very similar inhomogeneous $SL(2,\ints)$ invariant equations arise in the study of so-called modular graph functions~\cite{DHoker:2015gmr,DHoker:2015wxz,Basu:2016kli,Kleinschmidt:2017ege} and we shall also study these cases.

The structure of the inhomogeneous Laplace equation~\eqref{eq:D6R4} indicates that it will not be solved by an automorphic function of the standard type as these functions are required to be finite under all $E_{11-D}$-invariant differential operators, see for instance~\cite{Fleig:2015vky}. The generalised class of $E_{11-D}(\ints)$-invariant functions on $E_{11-D}$ to which the solution $\mathcal{E}_{(0,1)}^{(D)}$ belongs has not been identified abstractly. In this article, we will not attempt to define this class fully but rather present a method for solving equations of the type~\eqref{eq:D6R4} using a Poincar\'e series ansatz. 

Our method is inspired by the recent explicit solution of~\eqref{eq:D6R4} that was presented by Green, Miller and Vanhove in~\cite{Green:2014yxa} for the case of $D=10$ type IIB string theory. In this case one has the Cremmer--Julia group $SL(2,\reals)$ and U-duality group $SL(2,\ints)$ and they performed a Fourier expansion of the equation, using $SL(2,\ints)$-invariance and the known source function $\mathcal{E}_{(0,0)}^{(10)}$ that appears on the right-hand side of~\eqref{eq:D6R4}. The solution to the homogeneous equation is fixed by consistency with string perturbation theory as a boundary condition. In an appendix of~\cite{Green:2014yxa}, the authors rewrite their solution as a Poincar\'e series. The starting point of our approach is that the `seed' of the Poincar\'e series solves a much simpler Laplace equation than~\eqref{eq:D6R4}. This simpler equation is, however, still inhomogeneous. We present solutions to this equation. The hard part is now to impose the correct boundary conditions and to extract the Fourier expansion. Our method is explained in detail in section~\ref{sec:PSansatz}.

The correction term $D^6R^4$ has attracted a fair amount of attention recently~\cite{Basu:2014hsa,Pioline:2015yea,Bossard:2015uga}, in particular in connection to the Kawazumi--Zhang invariant of genus two Riemann surfaces~\cite{DHoker:2013eea,DHoker:2014gfa,Pioline:2015qha,Pioline:2015nfa}. In the supergravity limit, the corresponding expressions resemble two-loop field theory amplitudes~\cite{Pioline:2015yea}. This agrees nicely with independent expressions for (parts of) $\mathcal{E}_{(0,1)}^{(D)}$ that were recently found using a two-loop calculation in exceptional field theory with manifest $E_{11-D}$ invariance~\cite{Bossard:2015foa}. Other approaches to solving the inhomogeneous differential equation include the original approach followed in~\cite{Green:2005ba,Green:2014yxa} based on Fourier decomposing the equation, but one can also try to tackle the equation by using spectral methods~\cite{Green:2014yxa,Logan:2018} and there might also be a connection to automorphic distributions~\cite{Green:2014yxa}. Our present work is complementary to these results.

This article is structured as follows. In section~\ref{sec:PSansatz}, we present the new method for solving the inhomogeneous Laplace equation~\eqref{eq:D6R4} based on a Poincar\'e series ansatz. The method is used in section~\ref{sec:sols} to construct the function $\mathcal{E}_{(0,1)}^{(D)}$ in various dimensions while we consider an example from modular graph functions in section~\ref{sec:MGF}. Section~\ref{sec:concl} discusses open problems arising from our analysis.

\section{Poincar\'e series ansatz}
\label{sec:PSansatz}

After first reviewing briefly Eisenstein series and their properties, we expose our ansatz and how it can be used to simplify the inhomogeneous Laplace equation.

\subsection{A brief reminder of Eisenstein series}

The homogeneous Laplace equations in~\eqref{eq:Lapl} for $\mathcal{E}_{(0,0)}^{(D)}$ and $\mathcal{E}_{(1,0)}^{(D)}$ can be solved in terms of (combinations of) Eisenstein series $E(\lambda,g)$ on the symmetric space $E_{11-D}/K(E_{11-D})$. Such Eisenstein series are invariant under $E_{11-D}(\ints)$, can be parametrised by a (complex) weight $\lambda$ of the Lie algebra and satisfy the Laplace equation
\begin{align}
\left( \Delta  - \frac12(\lambda^2-\rho^2)\right) E(\lambda,g) = 0,
\end{align}
where $\rho$ is the Weyl vector of the Lie algebra and the norm-squares are calculated using the Killing metric, normalised to $2$ on real roots. Eisenstein series can be written as\footnote{This expression is absolutely convergent for large enough real parts of $\lambda$ (with respect to all simple roots) and can be analytically continued almost everywhere by functional equations~\cite{Langlands}.}
\begin{align}
\label{eq:eisen}
E(\lambda,g) = \sum_{\gamma \in B(\ints) \backslash G(\ints)} e^{\langle \lambda+\rho | H(\gamma g)\rangle},
\end{align}
where $H : E_{11-D} \to \mf{a}_{11-D}$ is a map from the group to the Cartan subalgebra $\mf{a}_{11-D}$ of the Lie algebra. It is given by the logarithm of the diagonal torus component $a$ in the Iwasawa decomposition $g=nak$ of a group element $g\in E_{11-D}$, where $n$ is in the maximal (upper) unipotent and $k\in K(E_{11-D})$. As the $k$-component of $g$ does not enter in the above definition, the Eisenstein series~\eqref{eq:eisen} descends to a function on the symmetric space $E_{11-D}/K(E_{11-D})$. In mathematical terms, it is spherical. Invariance of $E(\lambda,g)$ under $E_{11-D}(\ints)$ is manifest in~\eqref{eq:eisen} as it is an orbit sum; the quotient by $B(\ints)=N(\ints)A(\ints)$ is necessary as one has $H(\gamma g)=H(g)$ for  $\gamma\in B(\ints)$.

We can also think of the summand as arising from a $B(\ints)$-invariant  \textit{character} 
\begin{align}
\chi_\lambda :  B(\ints)\backslash B \to \cx^\times ,\quad \chi_\lambda(na) = e^{\langle \lambda+\rho | H( na)\rangle}
\end{align}
on the Borel subgroup $B$ that has been extended trivially to all of $E_{11-D}=B K(E_{11-D})$. The character property means that
\begin{align}
\chi_\lambda (b b') = \chi_\lambda(b) \chi_\lambda(b')
\end{align}
for any $b,b'\in B$.

\subsection{$R^4$ and $D^4R^4$ curvature corrections}
\label{sec:R4D4R4}

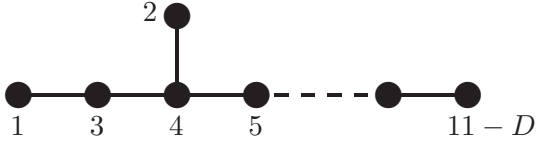
\begin{figure}
\centering
\begin{picture}(200,50)
\thicklines
\multiput(10,10)(30,0){4}{\circle*{10}}
\multiput(97,10)(10,0){5}{\line(1,0){5}}
\put(10,10){\line(1,0){90}}
\put(150,10){\line(1,0){30}}
\put(150,10){\circle*{10}}
\put(180,10){\circle*{10}}
\put(70,40){\circle*{10}}
\put(70,10){\line(0,1){30}}
\put(7,-5){$1$}
\put(57,38){$2$}
\put(37,-5){$3$}
\put(67,-5){$4$}
\put(97,-5){$5$}
\put(172,-5){$11-D$}
\end{picture}
\caption{\small Dynkin diagram of $E_{11-D}$ with (Bourbaki) labelling of nodes used in the text. \label{fig:dynk}}
\end{figure}

The functions $\mathcal{E}_{(0,0)}^{(D)}$ and $\mathcal{E}_{(1,0)}^{(D)}$ that appear for the $R^4$ and $D^4R^4$ curvature corrections, respectively, satisfy the homogeneous Laplace equations of the first two lines in~\eqref{eq:Lapl} unless there is a Kronecker source term. In the generic cases without Kronecker source term, the solutions are given by Eisenstein series of $E_{11-D}$ with weights given as in the following table.
\begin{center}
\begin{tabular}{c|c|c|c}
function & curvature term & weight $\lambda$ & (automorphic) representation\\[2mm]\hline&&&\\[-4mm]
$\mathcal{E}_{(0,0)}^{(D)}=2\zeta(3) E(\lambda,g)$ & $R^4$ & $\lambda =3\Lambda_1-\rho$ $(s=\tfrac32)$& minimal\\[2mm]
$\mathcal{E}_{(1,0)}^{(D)}=\zeta(5) E(\lambda,g)$ & $D^4R^4$ & $\lambda =5\Lambda_1-\rho$  $(s=\tfrac52)$& next-to-minimal
\end{tabular}
\end{center}
Here, the weight $\Lambda_1$ refers to the fundamental weight associated with node $1$ of the Dynkin diagram of $E_{11-D}$ as given in figure~\ref{fig:dynk}. The table lists the relevant weights for the Eisenstein series.\footnote{For $D<9$, the Eisenstein series for these weights have to be obtained by analytic continuation from a functionally related convergent expression of the form~\eqref{eq:eisen}.} In the last column, we also show which representations of $E_{11-D}$ the automorphic functions belong to. The above answers have passed many consistency checks~\cite{Green:2010wi,Pioline:2010kb,Green:2010kv} and were also found by direct exceptional field theory calculations~\cite{Bossard:2015foa}. 

The quasi-characters $\chi_\lambda$ for the particular weight $\lambda=2s\Lambda_1-\rho$ actually have a larger invariance than $B(\ints)$: Also the whole T-duality group $SO(10-D,10-D;\ints)$ (associated with the nodes $2,\ldots,11-D$ in the diagram) leaves $\chi_\lambda$ invariant. Combined with $B(\ints)$ one obtains an invariance under the maximal parabolic subgroup $P_1(\ints)$ and the Eisenstein series can also be called a maximal parabolic Eisenstein series~\cite{Green:2010wi}.

\subsection{Inhomogeneous Laplace equation and ansatz}
\label{sec:PA}

We reproduce the inhomogeneous equation~\eqref{eq:D6R4} for the $D^6R^4$ correction for convenience:
\begin{align}
\label{eq:IL}
\left(\Delta - \frac{6(14-D)(D-6)}{D-2} \right) \mathcal{E}^{(D)}_{(0,1)} = -\left(\mathcal{E}_{(0,0)}^{(D)}\right)^2+40\zeta(3)\delta_{D,6}+\frac{55}{3}\mathcal{E}_{(0,0)}^{(5)} \delta_{D,5}+\frac{85}{2\pi}\mathcal{E}_{(1,0)}^{(4)}\delta_{D,4}.
\end{align}
The Kronecker delta source terms in dimensions $D=4,5,6$ are related to supergravity divergences (also in form factors). If one works in a dimension $D$ where these do not arise, the equation is of the form
\begin{align}
\label{eq:IL2}
\left(\Delta-\mu\right) \varphi(g) = - (E(\lambda,g))^2,
\end{align}
where $E(\lambda,g)$ is an Eisenstein series that is given by a coset sum as in~\eqref{eq:eisen}. This equation has to be supplemented by appropriate boundary conditions near cusps of $G/K$. In string theory language this includes for example consistency with string perturbation theory~\cite{Green:2014yxa}. More generally, we could also allow for equations involving a polynomial in Eisenstein series on the right-hand side and this happens for higher derivative terms~\cite{Green:2008bf} or modular graph functions~\cite{DHoker:2015gmr,Kleinschmidt:2017ege}.

The strategy in~\cite{Green:2014yxa} for solving the equation~\eqref{eq:IL2} was to (double) Fourier expand both sides of the equation and then obtain simpler equations for the Fourier coefficients. A standard Fourier expansion of the right-hand side would require computing the convolution of the Fourier expansions of the product of two Eisenstein series, something that has not been carried out in full. The double Fourier expansion consisted in not doing the convolution sum but rather looking at each summand individually. In appendix~A of~\cite{Green:2014yxa}, the authors rewrite the doubly Fourier expanded solution in terms of a Poincar\'e series. This has served as inspiration for our strategy to directly work with a Poincar\'e series ansatz.

Our strategy  consists in making the ansatz that $\varphi(g)$ can be written as a coset sum 
\begin{align}
\label{eq:PA}
\varphi(g) = \sum_{\gamma \in B(\ints)\backslash G(\ints)}  \sigma(\gamma g),
\end{align}
with the `seed'
\begin{align}
\sigma : G \to \cx^\times 
\end{align}
that has the property that
\begin{align}
\sigma (b_\ints g k) = \sigma(g)
\end{align}
for all $b_\ints \in B(\ints)$ and $k\in K(E_{11-D})$. In other words, the seed $\sigma$ can be thought of as a function on the Borel subgroup $B$ that is trivial on the discrete subgroup $B(\ints)$ and extended trivially to all of $G$ by right $K$-invariance. Importantly, we do \textit{not} assume, however, that $\sigma$ is a character on $B$. Therefore, the Poincar\'e sum $\varphi(g)$ in~\eqref{eq:PA} is in general not an Eisenstein series.\footnote{The differential equation~\eqref{eq:IL2} for $\varphi$ also implies that $\varphi$ is not an automorphic function in the standard sense~\cite{Fleig:2015vky}. Standard automorphic functions are what is called $\mathcal{Z}(\mf{g})$-finite, where $\mathcal{Z}(\mf{g})$ is the center of the universal enveloping algebra of the Lie algebra $\mf{g}$ of $E_{11-D}$. This center is generated by the Casimir operators that translate into invariant differential operators. Because of the source $(E(\lambda,g))^2$ in~\eqref{eq:IL2} the system is not finite under the action of the Laplace operator.} Moreover, the ansatz entails that we would like the Poincar\'e sum to be absolutely convergent. As we shall see in the examples below, this is not always obvious to achieve.

With the ansatz~\eqref{eq:PA} and the expression~\eqref{eq:eisen} one can find solutions of the inhomogeneous Laplace equation~\eqref{eq:IL2} if one has a solution of the equation
\begin{align}
\label{eq:fold}
\left( \Delta - \mu \right) \sigma(g) = - \chi_{\lambda}(g) E(\lambda,g).
\end{align}
In writing this equation, we have `folded' the coset sum in one of the Eisenstein factors on the right-hand side. On the left-hand side one also uses the fact that the differential operator is invariant under the group $E_{11-D}$.

Since both $\sigma(g)$ and $\chi_\lambda(g)$ are left-invariant under the discrete Borel group $B(\ints)$, this group can be further used to Fourier expand equation~\eqref{eq:fold}. Let $U\subset B$ be a unipotent subgroup of the Borel group $B$. For example, we could choose $U=N$, the maximal unipotent subgroup generated by all positive root spaces, but this is not necessary and other choices might be more convenient in some cases. Let $\psi: U(\ints) \backslash U \to U(1)$ be a unitary character on $U$, i.e.\ for abelian $U$ it is of the form\footnote{For non-abelian $U$, one only has to include the generators associated with the abelianisation $[U,U]\backslash U$. The Fourier expansion will then be incomplete and has to be refined using also non-abelian Fourier coefficients~\cite{Bump,Pioline:2009qt,Fleig:2015vky}.}
\begin{align}
\psi(u) = \psi\left( \exp\left( \sum_{k=1}^{\dim U} x_k E_k \right) \right) = \exp\left( 2\pi i \sum_{k=1}^{\dim U} m_k x_k\right),
\end{align}
where $E_k$ are the suitably normalised nilpotent generators of $U$ and $m_k\in \ints$ are the mode numbers of the character $\psi$ that we will also refer to as instanton numbers or instanton charges. 

The Fourier coefficient of any function $f(g)$ that is left-invariant under $U(\ints)$ is then defined by
\begin{align}
f_\psi (g) = \int\limits_{U(\ints)\backslash U} f(ug) \overline{\psi(u)} du,
\end{align}
where the integration is over a single period and the Haar measure is normalised such that $U(\ints)\backslash U$ has unit volume. Fourier coefficients satisfy $f_\psi(ug) = \psi(u) f_\psi(g)$ for any $u\in U$. For abelian $U$ one has the Fourier expansion of the function $f$ given by
\begin{align}
f(g) = \sum_{\psi} f_\psi(g).
\end{align}
The sum is over all unitary characters, i.e.\ over all choices $m_k\in \ints$. The trivial case $\psi=1$ represents the constant term of $f$ with respect to $U$; in physics applications this comprises typically all perturbative contributions in some modulus.

Taking Fourier coefficients of equation~\eqref{eq:fold} one then obtains
\begin{align}
\left(\Delta-\mu\right) \sigma_\psi(g) = - \left(\chi_\lambda E\right)_\psi(g).
\end{align}
For the product right-hand side one needs to take the convolution of the Fourier expansions of the factors. Due to the Poincar\'e series ansatz~\eqref{eq:PA} the convolution has become trivial since the character $\chi_\lambda$ only has a zero mode in its Fourier expansion. Therefore one obtains
\begin{align}
\label{eq:PAF}
\left(\Delta-\mu\right) \sigma_\psi(g) = - \chi_\lambda(g)  E_\psi(\lambda,g).
\end{align}
If the Fourier expansion of $E(\lambda,g)$ is known (as it is for the $R^4$ cases in string theory), one can explicitly write down this equation and try solve it for each mode $\psi$, separately. This is the strategy we will follow in the following sections.

Besides convergence, one tricky point is the fate of the boundary conditions when making the Poincar\'e series ansatz. The inhomogeneous equation~\eqref{eq:PAF} will allow also require solutions to the homogeneous equation and the particular combination that arises for the wanted solution $\sigma$ is fixed by boundary conditions. However, these boundary conditions are originally stated for the Poincar\'e sum $\varphi$ in~\eqref{eq:PA}. We do not currently know how to generally translate these conditions directly into conditions for $\sigma$ and so in principle one has to perform the Poincar\'e sum to select the right solution. In the cases below we will basically follow this logic for fixing the homogeneous solution.

\section{$D^6R^4$ solutions in various dimensions}
\label{sec:sols}

in this section we implement the strategy just outlined to determine the $D^6R^4$ coefficient function in $D=10$ and $D=7$ space-time dimensions.

\subsection{Type IIB in $D=10$}
\label{sec:10D}

In the case of type IIB in $D=10$, the differential equation on the Poincar\'e upper half plane $\UHP=\left\{ z=x+iy\in \cx \middle|\, y>0\right\}\cong SL(2,\reals)/SO(2)$ is~\cite{Green:2014yxa}
\begin{align}
\label{eq:D6R410}
\left( \Delta - 12\right) \mathcal{E}_{(0,1)} =  - \left(\mathcal{E}_{(0,0)}\right)^2.
\end{align}
The $R^4$ function $\mathcal{E}_{(0,0)}$ on $\UHP$ is given by
\begin{align}
\mathcal{E}_{(0,0)} (z) = 2\zeta(3) E_{3/2}(z),
\end{align}
with $E_{3/2}(z)$ being the $s=\frac32$ case of the non-holomorphic Eisenstein series
\begin{align}
\label{eq:Es}
E_s(z)= \sum_{\gamma\in B(\ints)\backslash SL(2,\ints)}\left[\mathrm{Im} (\gamma z)\right]^s =  y^{s} + \frac{\xi(2(1-s))}{\xi(2s)} y^{1-s} + \sum_{n\neq 0} \mathcal{F}_{n,s}(y) e^{2\pi i n x}
\end{align}
with the non-zero Fourier coefficients
\begin{align}
\mathcal{F}_{n,s}(y) = \frac{2}{\xi(2s)} y^{1/2} |n|^{s-1/2} \sigma_{1-2s}(|n|) K_{s-1/2} (2\pi |n| y)
\end{align}
and $\xi(s) = \pi^{-s/2} \Gamma(s/2) \zeta(s)$. An element $\gamma=\left(\begin{smallmatrix}a&b\\c&d\end{smallmatrix}\right)\in SL(2)$ acts on $z\in\UHP$ by $z\mapsto \gamma z = \frac{az+b}{cz+d}$. $K_t(y)$ in the above is a modified Bessel function of the second kind and $\sigma_t(k) = \sum_{d|k} d^t$ is the (positive) divisor sum for $k\in\ints_{>0}$. The Eisenstein series $E_s(z)$ is induced from the character $\chi_s(z)=y^s$. The Laplace operator on $\UHP$ is $\Delta= y^2 \left(\partial_x^2+\partial_y^2\right)$.

The physical interpretation of the modulus $z\in\UHP$ here is such that its real part $x$ corresponds to the RR axion of type IIB string theory while its imaginary part $y$ is the inverse string coupling $\gs^{-1}$. The Fourier expansion of $E_{3/2}(z)$ given in~\eqref{eq:Es} then contains two zero mode terms, $y^{3/2}=\gs^{-3/2}$ and $y^{-1/2}=\gs^{1/2}$. These correspond to the perturbative tree-level and one-loop contributions to the four-graviton scattering amplitude expressed in Einstein frame; the fact that there are no further zero modes is a strong perturbative non-renormalisation theorem~\cite{Green:1997tv}.

Making the Poincar\'e series ansatz~\eqref{eq:PA} for $\mathcal{E}_{(0,1)}(z)$, \textit{viz.}
\begin{align}
\mathcal{E}_{(0,1)}(z) = \sum_{\gamma\in B(\ints)\backslash SL(2,\ints)} \sigma(\gamma z)
\end{align}
the inhomogeneous Laplace equation~\eqref{eq:D6R410} leads to the `folded' equation
\begin{align}
\left(\Delta-12\right) \sigma(z) =- 4\zeta(3)^2 y^{3/2} E_{3/2}(z).
\end{align}
This equation corresponds to~\eqref{eq:fold} in the general discussion and we can further analyse it by Fourier expanding both sides. The abelian unipotent invariance group here is given by $U(\ints)=N(\ints) =\left\{\left(\begin{smallmatrix} 1&k\\&1\end{smallmatrix}\right) \,\middle|\, k \in \ints\right\}$. Writing the Fourier expansion of $\sigma(z)$ as
\begin{align}
\sigma(z) = c_0(y) + \sum_{n\neq 0} c_n(y) e^{2\pi i n x}
\end{align}
leads to the following two equations for the zero and non-zero Fourier modes
\begin{subequations}
\begin{align}
\label{eq:zero10}
(y^2\partial_y^2-12) c_0(y) &= -4\zeta(3)^2 y^3 -\frac43 \pi^2 \zeta(3) y ,&\\
\label{eq:nonZ10}
(y^2\partial_y^2-4\pi^2 n^2 y^2-12) c_n(y) &= -16\pi \zeta(3) y^2 |n| \sigma_{-2}(|n|) K_1(2\pi|n| y),&
\end{align}
\end{subequations}
where we have used the explicit form of the Fourier coefficients for $E_{3/2}(z)$ given in~\eqref{eq:Es}. These equations correspond to~\eqref{eq:PAF} in the general discussion.

\vspace{3mm}

The general solution to equation ~\eqref{eq:zero10} for the zero mode $c_0(y)$ is simple to obtain:
\begin{align}
\label{czero}
c_0(y) = \frac23 \zeta(3)^2 y^3 +\frac1{9} \pi^2\zeta(3) y  + \alpha y^{-3} + \beta y^4.
\end{align}
Here, $\alpha$ and $\beta$ are \textit{a priori} undetermined integration constants that have to be fixed by boundary conditions. We note that performing the Poincar\'e sum of any term of the form $y^s$ produces the Eisenstein series $E_s(z)$ with perturbative terms $y^s$ and $y^{1-s}$, cf.~\eqref{eq:Es}. Having either $\alpha$ or $\beta$ non-zero will therefore necessarily lead to the two zero mode terms $y^{-3}$ and $y^4$ in the summed $\mathcal{E}_{(0,1)}$. While the term $y^{-3}=\gs^3$ corresponds to a three-loop contribution (after changing to string frame), the term $y^4=\gs^{-4}$ would be a contribution of loop order $-1/2$, something that is incompatible with string perturbation theory. Since both homogeneous solutions parametrised by $\alpha$ and $\beta$ would lead to this inconsistent behaviour, we are led to set
\begin{align}
\label{eq:solZ10}
\alpha=\beta=0  
\quad\quad\Rightarrow\quad\quad
c_0(y) = \frac23 \zeta(3)^2 y^3 +\frac23 \zeta(2)\zeta(3) y.
\end{align}

\vspace{3mm}

The solution of equation~\eqref{eq:nonZ10} for the non-zero modes $c_n(y)$ is more complicated. The homogeneous equation can be recast in Bessel form and has only one solution that falls off at the weak coupling cusp $y=\gs^{-1}\to\infty$. For mode number $n$ it is given by $y^{1/2}K_{7/2} (2\pi |n| y)$ and has in fact a finite asymptotic expansion around weak coupling. A particular solution of~\eqref{eq:nonZ10} can be extracted from the analysis~\cite{Green:2014yxa}\footnote{In appendix~\ref{app:ps}, we present an algebraic formalism that also leads to this solution.} and combined with the homogeneous solution we obtain
\begin{align}
\label{eq:IIBnz}
c_n(y) &= \frac{8 \zeta (3) \sigma_{-2}(|n|)}{|n|}\left[\left(|n|y+\frac{10}{\pi ^2 y |n|}\right) K_0(2 \pi |n| y)
+ \left(\frac{6}{\pi}+\frac{10}{\pi ^3 y^2 |n|^2}\right) K_1(2 \pi  |n|y) \right]&\nn\\
&\quad + \alpha_n y^{1/2} K_{7/2} (2\pi |n| y),
\end{align}
where $\alpha_n$ is the integration constant associated with the homogeneous solution. Again, this integration constant has to be fixed from asymptotic considerations. In this case, we consider the strong coupling region $y\to 0$ where these instantonic contributions dominate.  By S-duality this region is also related to the perturbative regime $y\to\infty$. The absence of any singular terms in the limit $y\to 0$ determines the value of $\alpha_n$ to be
\begin{align}
\alpha_n = -\frac{128 \zeta (3) \sigma_{-2}(|n|)}{3 \pi  \sqrt{|n|}},
\end{align}
leading to the final expression after some rearrangements
\begin{align}
\label{eq:solNZ10}
c_n(y) &= 8 \zeta (3) \sigma_{-2}(|n|) y\bigg[\left(1+\frac{40}{(2\pi  |n|y)^2}\right) K_0(2 \pi |n| y)
+ \left(\frac{12}{2\pi |n| y}+\frac{80}{(2\pi  |n|y)^3}\right) K_1(2 \pi  |n|y) &\nn\\
&\quad\quad\quad\quad\quad\quad\quad\quad\quad -\frac{16}{3\pi (|n|y)^{1/2}} K_{7/2} (2\pi |n| y)\bigg].
\end{align}
This way $c_n(y)$ is of order $y$ as $y\to 0$ and does not contain any singular terms.\footnote{This condition appears slightly stronger than the requirement $O(y^{-2})$ as $y\to0$ for the summed solution $\mathcal{E}_{(0,1)}(z)$  that was found by S-duality in Lemma 2.9 in~\cite{Green:2014yxa}.} The resulting expression agrees precisely with the result found in~\cite{Green:2014yxa} but now obtained directly from a Poincar\'e series ansatz.

We note that the seed function $\sigma(z) = c_0(y) + \sum_{n\neq 0} c_n(y)e^{2\pi i n x}$ depends non-trivially on both Borel coordinates $x$ and $y$ and in particular is \textit{not} a character on $B$. This leads to complications when carrying out the Poincar\'e sum. In fact, the full expression is not known. 
Let us make a few comments about convergence of the Poincar\'e sum over the seed we just determined. If the sum were absolutely convergent one could perform the sum over all the terms separately. This cannot be true as it is well known that the Poincar\'e sum for the linear term in $y$ in $c_0(y)$ does not converge: it represents the limiting value of the non-holomorphic Eisenstein series for $SL(2$). It is therefore desirable to introduce a regularised version of the solution depending on a parameter with a limit that is related to the above solution.\footnote{In terms of the Eisenstein series $E_s(z)$ induced by the character $y^s$ one obtains the so-called (first) Kronecker limit formula~\cite{Fleig:2015vky} for $s\to 1$.} In appendix~\ref{app:reg}, we present a regularised version of the equation and the solution where this problem does not arise.

The physical content of the solution is obtained by performing a Fourier expansion of $\mathcal{E}_{(0,1)}(z)=\sum_{\gamma} \sigma(\gamma z)$ with respect to the periodic real part $x$ that represents the Ramond--Ramond axion in type IIB theory. The zero mode piece in $\mathcal{E}_{(0,1)}(z)$  corresponds to the perturbative terms in $\gs$, together with non-perturbative contributions of vanishing net instanton charge. These should be interpreted as instanton-anti-instanton contributions and they are exponentially suppressed by $e^{-4\pi |n| y}$~\cite{Green:2005ba}. We have not managed to derive the Fourier expansion from the Poincar\'e sum form and here merely quote the result for the perturbative zero modes obtained in~\cite{Green:2005ba,Green:2014yxa}\footnote{The exponentially suppressed terms in the zero and non-zero modes are not known explicitly.}
\begin{align}
\label{eq:IIBpert}
\mathcal{E}_{(0,1)}^{(10B,\mathrm{pert.})} = \frac23 \zeta(3)^2 y^3 + \frac43 \zeta(2)\zeta(3) y + \frac{4\zeta(4)}{y} + \frac{4\zeta(6)}{27y^3}\,.
\end{align}
The perturbative terms thus obtained correspond to contributions from tree-level up to three-loops. The numerical values obtained by using $SL(2,\ints)$ invariance and the differential equation have been confirmed by direct string theory calculations~\cite{Green:2005ba,Gomez:2013sla,DHoker:2013eea}. 

\subsection{Toroidal compactification to $D=7$}

In the case of $D = 7$, the U-duality group is $SL(5,\ints)$ and the moduli space is $SL(5)/SO(5)$. Equation \eqref{eq:D6R4} becomes
\begin{equation}
    \label{eq:SL5D6R4}
    \left( \Delta  - \frac{42}{5} \right) \mathcal{E}^{(7)}_{(0, 1)} = -\left( \mathcal{E}^{(7)}_{(0, 0)} \right)^{2}\,,
\end{equation}
where the $R^4$-function $\mathcal{E}^{(7)}_{(0, 0)}(g) = 2\zeta(3) E(3 \Lambda_1 - \rho, g)$ satisfies
\begin{equation}
    \left( \Delta  + \frac{12}{5} \right) \mathcal{E}^{(7)}_{(0, 0)} = 0\,. 
\end{equation}
We will analyse equation~\eqref{eq:SL5D6R4} over the mirabolic $P_1$, i.e.\ the maximal parabolic subgroup associated with node $1$. This is the parabolic that is associated with the string perturbation theory expansion; the Levi is $SL(4)\times GL(1)\cong SO(3,3)\times \reals_+$ (locally) giving the T-duality moduli space of the string theory three-torus and the string coupling. Explicitly, we parametrise the group element $g\in SL(5)$ as
\begin{equation}
\label{eq:7string}
    g = ulk = 
    \left(
    \begin{matrix}
        1 & Q \\
        0 & \I
    \end{matrix}
    \right)
    \left(
    \begin{matrix}
        r^{4/5} & 0 \\
        0 & r^{-1/5}e_{4}
    \end{matrix}
    \right)k
\end{equation}
where $Q$ is a four-component row vector and $e_4$ is a general element of $SL(4)$ and $k\in SO(5)$. The variable $r$ equals the inverse string coupling\footnote{We parametrise the string coupling in $D$ dimensions $g_D$ such that different orders in perturbation theory differ by $g_D^2$; this is different from the convention used in~\cite{Green:2010kv}.} in $D=7$, $Q$ corresponds to the four axions that BPS-instantons couple to\footnote{In type IIA language, there are three D$0$-instantons (wrapping one of the three cycles of $T^3$) and one D$2$-instanton (wrapping the full torus). In type IIB language, there is one (point-like) D$(-1)$-instanton and three  D$1$-instantons (wrapping two out of three cycles).} and $e_4\in SL(4)\cong SO(3,3)$ is associated with the moduli space of $T^3$.
The Laplacian $\Delta = \Delta_{SL(5)}$ on $SL(5)/SO(5)$ decomposes in these coordinates  as
\begin{equation}
    \Delta_{SL(5)} = \frac{5}{8}r^2 \partial_r^2 - \frac{15}{8}r \partial_r + r^2 ||e_4^{-1} \partial_{Q}||^2 + \Delta_{SL(4)}\,.
\end{equation}

\subsubsection{Perturbative terms in the string coupling}

Before solving the inhomogeneous equation~\eqref{eq:SL5D6R4} using a Poincar\'e sum, we first consider the perturbative pieces similar to~\eqref{eq:IIBpert} by considering the zero Fourier modes in an expansion of $\mathcal{E}_{(0,1)}^{(7)}$ in the decomposition~\eqref{eq:7string}. 
A similar analysis can be found in~\cite{Green:2010wi}. Making the ansatz for the perturbative terms up to three loops with $F_h$ denoting the $h$-loop perturbative piece as\footnote{The overall pre-factor comes from relating the string scale to the Planck scale in $D=7$ space-time dimensions.}
\begin{align}
\mathcal{E}_{(0,1)}^{(7,\mathrm{pert.})} &= r^{14/5} \left( r^2 F_0 + r^0 F_1 + r^{-2} F_2 +r^{-4} F_3\right)\nn\\
 &= r^{24/5} F_0 + r^{14/5} F_1 + r^{4/5} F_2 +r^{-6/5}F_3
\end{align}
leads to the four equations 
\begin{subequations}
\label{eq:Fhall}
\begin{align}
\left(\Delta_{SL(4)}-6\right) F_0  &= -4\zeta(3)^2\,,\\
\label{eq:F2}
\left(\Delta_{SL(4)}-\frac{21}2\right) F_1  &= -16\zeta(2)\zeta(3) E^{SL(4)}(2\Lambda_1-\rho,e_4)\,,\\
\left(\Delta_{SL(4)}-10\right) F_2  &= -16\zeta(2)^2 E^{SL(4)}(2\Lambda_1-\rho,e_4)^2\,,\\
\left(\Delta_{SL(4)}-\frac92\right) F_3  &=0\,.
\end{align}
\end{subequations}
The equations not involving a squared source on the right-hand side are solved by
\begin{subequations}
\begin{align}
F_0 &=\frac23 \zeta(3)^2\,,\\
F_1 &=\frac43 \zeta(2) \zeta(3) E^{SL(4)}(2\Lambda_1-\rho,e_4)+\frac{5\pi}{756} \zeta(7) E^{SL(4)}(7\Lambda_2-\rho,e_4)\,,\\
F_3 &= 4\zeta(6) \left( E^{SL(4)}(6\Lambda_1-\rho,e_4) + E^{SL(4)}(6 \Lambda_3-\rho,e_4)\right)\,.
\end{align}
\end{subequations}
The constant tree-level contribution follows by expanding the known amplitude which also rules out any contributions from the kernel of the differential operator. The one-loop piece is a theta lift of the Narain partition function~\cite{Green:2008uj,Angelantonj:2011br}
\begin{align}
F_1 = \frac{\pi}{3} \int\limits_{SL(2,\ints)\backslash \mathbb{H}} \frac{d^2\tau}{\tau_2^2} \Gamma_{(3,3)}(e_4) \big[\xi(3) E_3(\tau) + \zeta(3)\big]\,,
\end{align}
where $\tau$ is the complex structure of the string one-loop torus and the combination $\xi(3) E_3(\tau) + \zeta(3)$  can also be obtained using modular graph functions~\cite{DHoker:2015gmr}. The second term in~\eqref{eq:F2} is in the kernel of the differential operator but is needed for obtaining the right decompactification limit.  The three-loop equation is homogeneous and the particular combination of the homogeneous solutions is fixed by having the right decompactification limit consistent with~\eqref{eq:IIBpert}.  It is also given by the genus-three theta lift of the constant function~\cite{Green:2010wi}. 
The two-loop term is as always the hardest as it satisfies a similar inhomogeneous equation to the original function. It is connected to an integral over the Kawazumi--Zhang invariant~\cite{DHoker:2013eea,DHoker:2014gfa} and constrained by having the right decompactification limit.

\subsubsection{Solution using a Poincar\'e sum}

Let us assume that $\mathcal{E}_{(0, 1)}^{(7)}$ is a Poincar\'e series with respect to the same maximal parabolic $P_1$ as in~\eqref{eq:PA}, i.e.\
\begin{equation}
\label{eq:PSA7}
    \mathcal{E}_{(0, 1)}^{(7)}(g) = \sum_{\gamma \in P_1(\ints) \backslash SL(5, \ints)}\sigma(\gamma\cdot g)\,.
\end{equation}
We note that this particular type of parabolic coset sum is an assumption. Our motivation for this choice is that is adapted to a string perturbation formulation of the solution, i.e.\ the seed will be decomposed into terms at fixed order in string perturbation theory plus $SO(3,3,\ints)$ T-duality invariant functions coupled to instantons. Of course, this ansatz for the seed gets spread out by the $SL(5,\ints)$ orbit sum into a more complicated U-duality invariant function. We shall find a solution to the differential equation with our parabolic ansatz; it is likely that other forms using other parabolics sums exist and it would be very interesting to study their relation and functional equations.

With the assumption of the parabolic Poincar\'e series~\eqref{eq:PSA7}, the Laplace equation \eqref{eq:SL5D6R4} then unfolds into
\begin{equation}
    \label{eq:unfoldedSL5D6R4}
    \left( \Delta  - \frac{42}{5} \right) \sigma(g) = -4\zeta(3)^2 r^{12/5} E(3\Lambda_1 - \rho, g)
    .
\end{equation}
Since we assume $\sigma$ to be left $P_1(\ints)$-invariant, it can be expressed as the Fourier series 
\begin{equation}
\label{eq:sigma5FE}
    \sigma(g) = \sum_{N \in \ints^{4}} c_{N}(r, e_4) e^{2\pi i Q  N}\,.
\end{equation}
As the unipotent of a maximal parabolic subgroup of $SL(n)$ is abelian, this expression captures the whole of $\sigma$ without need for non-abelian coefficients. Here,  $N$ is a four-component column-vector. The Eisenstein series on the right-hand side of \eqref{eq:unfoldedSL5D6R4} can also be written as a Fourier series over the unipotent as
\begin{equation}
    E(3\Lambda_1 - \rho, g) = \sum_{N \in \ints^4} f_{N}(r, e_4) e^{2\pi i Q  N}
\end{equation}
with
\begin{subequations}
\begin{align}
\label{eq:R4Fn}
    f_{N}(r, e_4) ={}& \frac{2}{\zeta(3)}r^{7/5} \sigma_2(k) \frac{K_1(2\pi r ||e_4^{-1}N||)}{||e_4^{-1} N||} \quad \text{for $N \neq 0$ and}  \\
    f_{0}(r, e_4) ={}& r^{12/5} + \frac{2 \zeta(2)}{\zeta(3)}r^{2/5} E^{SL(4)}(2\Lambda_1 - \rho, e_4)\,.
\end{align}
\end{subequations}
One has $\Delta_{SL(4)}E^{SL(4)}(2\Lambda_1 - \rho, e_4) = -\frac{3}{2}E^{SL(4)}(2\Lambda_1 - \rho, e_4)$ and $k=\gcd(N)$. 
The notation here is such that $\Lambda_1$ for $SL(4)$ denotes the node adjacent to the one used defining the string perturbation limit for $SL(5)$ and thus is one of the outer notes of $SL(4)$. (From the point of view of $SO(3,3)$, this is a spinor node.) We now obtain differential equations for the Fourier coefficients $c_{N}$ of $\sigma$ in~\eqref{eq:sigma5FE}.

\vspace{3mm}


For the zero mode $c_0$ we have the equation
\begin{equation}
    \left( \frac{5}{8}r^2 \partial_r^2 - \frac{15}{8}r \partial_r + \Delta_{SL(4)} - \frac{42}{5} \right) c_0(r, e_4) = -4\zeta(3)^2 r^{24/5} - 8 \zeta(2)\zeta(3) r^{14/5} E^{SL(4)}(2\Lambda_1 - \rho, e_4)
    .
\end{equation}
Starting with the homogeneous equation, one can make a separated ansatz of the form $c_0^{(\mathrm{h})}(r, e_4) = r^\alpha F_\alpha(e_4)$. This leads to
\begin{align}
\left(\Delta_{SL(4)} +\frac58 \alpha^2 -\frac52 \alpha -\frac{42}5\right) F_\alpha(e_4)  = 0\,.
\end{align}
In order for this to produce terms consistent with string perturbation theory only the values $\alpha\in \{\frac{24}5,\frac{14}5,\frac{4}5,-\frac65\}$ are allowed, leading to the eigenvalues $\{6,\frac{21}2,10,\frac92\}$ as in~\eqref{eq:Fhall}. The remaining equation for $F_\alpha(e_4)$ is then solved by appropriate $SL(4)$ Eisenstein series.

Looking for a particular solution of the form $\alpha_1 r^{24/5} + \alpha_2 r^{14/5} E^{SL(4)}(2\Lambda_1 - \rho, e_4)$ we are led to the particular solution
\begin{equation}
    c_0^{(\mathrm{p})}(r, e_4) = \frac23\zeta(3)^2 r^{24/5} + \frac23 \zeta(2)\zeta(3) r^{14/5}E(2\Lambda_1 - \rho, e_4)
    .
\end{equation}
The full solution for the zero mode is now $c_0 = c_0^{(\mathrm{h})} + c_0^{(\mathrm{p})}$. Comparison with the IIB case suggests that one should choose the homogeneous solutions $ c_0^{(\mathrm{h})} = \frac{5\pi}{1512} \zeta(7) r^{14/5} E^{SL(4)}(7\Lambda_2-\rho,e_4)$ but without computing the Fourier expansion of the Poincar\'e sum~\eqref{eq:PSA7} and comparing with~\eqref{eq:Fhall} we cannot fix the homogeneous term definitively.

\vspace{3mm}

For the non-zero modes $c_{N}(r,e_4)$, we have the equation
\begin{equation}
\left( \frac{5}{8}r^2\partial_r^2 - \frac{15}{8}r\partial_r - 4\pi^2 r^2 ||e_4^{-1}N||^2 + \Delta_{SL(4)} - \frac{42}{5} \right) c_{N} = -16\pi\zeta(3)\sigma_2(k) r^{19/5}\frac{K_1(2\pi r ||e_4^{-1}N||)}{||e_4^{-1} N||}\,.
    \label{eqn:nonzeromodesl5}
\end{equation}

We show in appendix~\ref{app:ps} that 
\begin{align}
c_N^{(\mathrm{p})} 
        = 32\pi^2\zeta(3)\sigma_2(k) r^{24/5} \bigg( \frac{K_0\left( 2\pi r ||e^{-1} N || \right)}{ \left( 2\pi r ||e^{-1} N|| \right)^2} + 12\frac{K_1\left( 2\pi r ||e^{-1} N || \right)}{\left( 2\pi r ||e^{-1} N|| \right)^3} + 40\frac{K_2\left( 2\pi r ||e^{-1} N || \right)}{\left( 2\pi r ||e^{-1} N|| \right)^4} \bigg) 
\end{align}
is a particular solution of this equation. The proof relies on writing out the $SL(4)$ Laplacian and using properties of the Bessel functions in a way similar to demonstrating~\eqref{eq:IIBnz}. As also explained in the appendix, there are solutions to the homogeneous equation given by (see~\eqref{eq:hom7})\footnote{We note that these homogeneous solutions appear not to correspond to the Fourier modes of an $SL(5)$ Eisenstein series associated with the minimal series on node $1$. Looking for such an Eisenstein series leads to irrational powers of $r$. There is, however, a well-known homogeneous solution given by $E^{SL(5)}(7\Lambda_2-\rho)$~\cite{Pioline:2015yea}.}
\begin{align}
c_N^{(\mathrm{h})} = r^{24/5} \frac{K_{7/2}(2\pi r ||e_4^{-1}N||)}{(2\pi r ||e_4^{-1} N||)^{5/2} } 
\quad\textrm{and}\quad
c_N^{(\mathrm{h})} = r^{16/5} \frac{K_{7/2}(2\pi r ||e_4^{-1}N||)}{(2\pi r ||e_4^{-1} N||)^{3/2} } \,.
\end{align}
If we impose the same constraint as for $SL(2)$, namely that the strong coupling behaviour $r\to 0$ is regular, this rules out the second solution and selects the combination
\begin{align}
c_N(r,e_4) &= 32\pi^2\zeta(3)\sigma_2(k) r^{24/5} \bigg[ \frac{K_0\left( 2\pi r ||e^{-1} N || \right)}{ \left( 2\pi r ||e^{-1} N|| \right)^2} + 12\frac{K_1\left( 2\pi r ||e^{-1} N || \right)}{\left( 2\pi r ||e^{-1} N|| \right)^3} + 40\frac{K_2\left( 2\pi r ||e^{-1} N || \right)}{\left( 2\pi r ||e^{-1} N|| \right)^4}  \nn\\
&\quad -\frac{2\sqrt{2}}{15\sqrt{\pi}}  \frac{K_{7/2}(2\pi r ||e_4^{-1}N||)}{(2\pi r ||e_4^{-1} N||)^{5/2} }\bigg]\,.
\end{align}

\section{Modular graph functions}
\label{sec:MGF}

Another family of automorphic functions satisfying inhomogeneous Laplace equation is provided by modular graph functions~\cite{DHoker:2015gmr}. These are functions that are invariant under $SL(2,\ints)$ and have an explicit lattice sum description. Moreover, they satisfy typically inhomogeneous Laplace equations. We note that for any Poincar\'e sum of the form $\varphi(z)=\sum_{\gamma\in B(\ints)\backslash SL(2,\ints)} \sigma(\gamma z)$, where $z=x+iy$ and the periodic seed has the expansion $\sigma(x+iy) = \sum_{n\in\ints} c_n(y) e^{2\pi in x}$,
the Fourier modes of $\varphi(z) = \sum_{n\in \ints} f_n(y) e^{2\pi i n x}$ are given by~\cite{Iwaniec,Fleig:2015vky}
\begin{align}
\label{eq:PSsumFE}
f_n(y) &= c_n(y) + \sum_{d>0} \sum_{m\in\ints} S(m,n;d) \int\limits_{\reals} \exp\left[-2\pi i \omega n -2\pi i m \frac{\omega}{d^2(\omega^2+y^2)}\right] c_m\left(\frac{y}{d^2(\omega^2+y^2)}\right)d\omega\,,
\end{align}
involving the Kloosterman sums
\begin{align}
S(m,n;d) = \sum_{q\in (\ints/d\ints)^\times} e^{2\pi i (qm+q^{-1}n)/d}\,.
\end{align}
This can be shown by writing out explicitly the coset sum. We shall try to apply this formalism to re-derive some results on modular graph functions.

As an example we consider the function $C_{3,1,1}(z)$ in the notation of~\cite{DHoker:2015gmr}. It can be defined explicitly from a multiple lattice sum as
\begin{align}
C_{3,1,1}(z) = \sum_{\substack{(m_1,n_1), (m_2,n_2)\in\ints^2\\ (m_i,n_i)\neq(0,0)\\(m_1+m_2,n_1+n_2)\neq (0,0)}} \frac{y^5}{\pi^5 |m_1z + n_1|^6 |m_2z+n_2|^2 |(m_1+m_2)z + (n_1+n_2)|^2}\,.
\end{align}
As shown in~\cite[Eq.~(3.19)]{DHoker:2015gmr} it satisfies the equation
\begin{align}
\label{eq:c311}
\left(\Delta-6\right) C_{3,1,1}(z) = \frac{86}{5}\ME_5 -4\ME_2\ME_3 + \frac{\zeta(5)}{10}\,,
\end{align}
where $\ME_s = 2\pi^{-s} \zeta(2s) E_s$ is the non-holomorphic Eisenstein series $E_s$ in a different normalisation.

We shall now forget that we have an explicit solution for $C_{3,1,1}(z)$ and try to construct one by solving the Laplace equation~\eqref{eq:c311} using a Poincar\'e ansatz. In order to treat the finite constant on the right-hand side, we replace it by an Eisenstein series $E_\epsilon$ and send $\epsilon\to 0$ at the end, using analytic continuation. Writing $C_{3,1,1}(z) = \sum_{\gamma\in B(\ints) \backslash SL(2,\ints)} \sigma(\gamma z)$ in Poincar\'e form, we have to solve the equation
\begin{align}
\left(\Delta-6\right) \sigma(z) = \frac{172\pi^5}{467\ 775} y^5 - \frac{8\pi^5}{42\ 525}y^3 E_2 + \frac{\zeta(5)}{10} y^{\epsilon}\,,
\end{align}
where we have explicitly written out the normalising factors and $y^\epsilon$ represents the regulator for the constant. In writing the equation we have chosen to `fold' $\ME_3$. The boundary condition for solving this equation is that the solution $f(z)$ should not have a term growing as $y^3$ when approaching $y\to\infty$.

Reducing the equation to Fourier modes $\sigma(z) =\sum_{n\in\ints}c_n(y)e^{2\pi i nx}$ leads to
\begin{align}
\left( y^2 \partial_y^2 - 6 \right) c_0(y) &=  \frac{4\pi^5}{22\ 275} y^5  -\frac{8\pi^2}{945} \zeta(3) y^{2}+ \frac{\zeta(5)}{10} y^{\epsilon}\,,\nn\\
\left( y^2 \partial_y^2 -4\pi^2n^2y^2- 6 \right) c_n(y) &= -\frac{32\pi^3}{945} |n|^{3/2} \sigma_{-3}(n)  y^{7/2} K_{3/2}(2\pi|n| y)\nn\\
&= -\frac{8\pi^2}{945} \sigma_{-3}(n) y^2 (1+2\pi|n| y) e^{-2\pi|n| y} \,.
\end{align}
A particular solution to these equations is given by
\begin{align}
c_0(y) &=  \frac{2\pi^5}{155\ 925}y^5  + \frac{2\pi^2\zeta(3)}{945} y^2 + \frac{\zeta(5)}{10 (\epsilon(\epsilon-1)-6)}y^\epsilon\,,\nn\\
c_n(y) &=  \frac{2\pi^2}{945} \sigma_{-3}(n) y^2 e^{-2\pi|n| y} \,.
\end{align}
(The homogeneous solutions correspond to the various Fourier modes of $E_2(z)$, but we will not require them here.)

Using the general formula for Fourier expansions of Poincar\'e sums one can now construct the zero mode of the $C_{3,1,1}(z)=\sum_{n\in \ints} f_n(y)e^{2\pi i n x}$ by computing
\begin{align}
f_0 (y) = c_0(y) + \sum_{d>0} \sum_{m\in\ints} S(m,0;d) \int_{\reals} \exp\left[-2\pi i m \frac{\omega}{d^2(\omega^2+y^2)}\right] c_m\left(\frac{y}{d^2(\omega^2+y^2)}\right)d\omega\,.
\end{align}
This expression can be evaluated for the present case as follows. We first rescale the integration variable and restrict to the terms with $m\neq 0$ in the sum. They are\allowdisplaybreaks{
\begin{align}
\label{eq:smneq0}
 &\quad y \sum_{d>0} \sum_{m\neq 0} S(m,0;d) \int_{\reals} \left[-2\pi i m y^{-1}d^{-2} \frac{t}{1+t^2}\right] c_m\left(\frac{y^{-1}d^{-2}}{1+t^2}\right)dt\nn\\
&= \frac{4\pi^2}{945} y^{-1}\sum_{m>0} \sum_{d>0} S(m,0;d) d^{-4}\sigma_{-3}(m)  \int_\reals (1+t^2)^{-2} \exp\left[-2\pi  m y^{-1} d^{-2} \frac{1+it}{1+t^2}\right]dt \nn\\
&= \frac{4\pi^2}{945} y^{-1}\sum_{m>0}\sum_{k \geq 0} \frac{1}{k!}\underbrace{\sum_{d>0} S(m,0;d) d^{-4-2k}}_{=\frac{\sigma_{-3-2k}(m)}{\zeta(4+2k)}} m^k \sigma_{-3}(m) (-2\pi y^{-1})^k \underbrace{\int_\reals (1+t^2)^{-k-2} (1+it)^k dt}_{= 2^{-k-2} (k+2) \pi}\nn\\
&= \frac{\pi^3}{945} y^{-1}\sum_{m>0} \sum_{k\geq 0} \frac{k+2}{k!}  \frac{\sigma_{-3}(m)\sigma_{-3-2k}(m) m^k}{\zeta(4+2k)} (-\pi y^{-1})^k\,.
\end{align}
As is evident from this expression the sum over $m$ is divergent and so the two remaining summations cannot be interchanged strictly. One can partly make sense of the sum over $m$ by analytically continuing the Ramanujan identity
\begin{align}
\sum_{m>0} \sigma_a(m) \sigma_b(m) m^{-s} = \frac{\zeta(s)\zeta(s-a)\zeta(s-b)\zeta(s-a-b)}{\zeta(2s-a-b)}
\end{align}
from the convergent region with large real part of $s$ to $s=-k$. This leads to
\begin{align}
\eqref{eq:smneq0}&= \frac{\pi^3}{945} y^{-1} \sum_{k\geq 0} \frac{k+2}{k!} \frac{\zeta(-k)\zeta(3-k)\zeta(k+3)\zeta(k+6)}{\zeta(4+2k)\zeta(6)} (-\pi y^{-1})^k\nn\\
&= -\frac{2 \zeta(3)^2}{21\pi} y^{-1} +\frac{7\zeta(7)}{16\pi^2}y^{-2} -\frac{\zeta(3)\zeta(5)}{2\pi^3 } y^{-3} + \frac{11\zeta(9)}{32\pi^{4}} y^{-4}\,.
\end{align}
The $k$-summation terminates due to the zeroes of the Riemann zeta function at negative even integers. The case $k=2$ requires taking a limit.}

To complete the calculation of the zero mode $f_0(y)$ we also have to take into account the contribution from $c_0(y)$ and the terms with $m=0$ in the general expression. These are just the usual constant terms for Eisenstein series~\cite{Fleig:2015vky}. This leads in total to
\begin{align}
f_0(y) &=  \frac{2\pi^5}{155\ 925}y^5  + \frac{2\pi^2\zeta(3)}{945} y^2 + \frac{\zeta(5)}{10 (\epsilon(\epsilon-1)-6)}y^\epsilon \nn\\
&\quad +  \frac{2\pi^5}{155\ 925} \frac{\xi(9)}{\xi(10)} y^{-4}  + \frac{2\pi^2\zeta(3)}{945} \frac{\xi(3)}{\xi(4)}y^{-1} + \frac{\zeta(5)}{10 (\epsilon(\epsilon-1)-6)} \frac{\xi(2\epsilon-1)}{\xi(2\epsilon)}y^{1-\epsilon}\nn\\
&\quad  -\frac{2 \zeta(3)^2}{21\pi} y^{-1} +\frac{7\zeta(7)}{16\pi^2}y^{-2} -\frac{\zeta(3)\zeta(5)}{2\pi^3 } y^{-3} + \frac{11\zeta(9)}{32\pi^{4}} y^{-4} \nn\\
&\stackrel{\epsilon\to 0 }{\to}\frac{2\pi^5}{155\ 925}y^5    + \frac{2\pi^2\zeta(3)}{945} y^2- \frac{\zeta(5)}{60}
+\frac{7\zeta(7)}{16\pi^2}y^{-2} -\frac{\zeta(3)\zeta(5)}{2\pi^3 } y^{-3}+\frac{43\zeta(9)}{64\pi^4} y^{-4}
\end{align}
This agrees with the Laurent polynomial stated in~\cite[Eq.~(6.2)]{DHoker:2015gmr} except for the constant term in $y^0$.\footnote{We note that this Laurent polynomial can be deduced by considering the zero mode of equation~\eqref{eq:c311} and applying the Rankin--Selberg method to fix the solution to the homogeneous solution.}  The reason for this appears to be that the constant term of $\ME_2\ME_3$ has a contribution to $y^0$ that is missed by the above construction and this seems to be a general feature that requires additional study. A further point that requires investigation is that our method of treating the divergent sum over $m$ seems to have lost all the exponentially suppressed terms of the form $O(e^{-y})$ in the zero mode while one would expect them from the known function $C_{3,1,1}(z)$. However, we note that the method produces also the right combination of solutions to the homogeneous equation found in~\cite{DHoker:2015gmr}.

\section{Discussion}
\label{sec:concl}

In the present paper, we have outlined a method for solving inhomogeneous automorphic differential equations of the type that appear in string theory in several places. This method relied on making an ansatz for the solution of a Poincar\'e sum form as in~\eqref{eq:PA}. The advantage of this method is that the resulting differential equation for the seed $\sigma$ of the Poincar\'e sum is less involved than the original equation and is solved in a Fourier expansion.  We exemplified this method for four examples, namely the $D^6R^4$ correction for ten-dimensional type IIB string theory and the $D^6R^4$ correction in $D=7$ space-time dimensions. The first example reproduced a known result from~\cite{Green:2005ba,Green:2014yxa} while the second example gave a new proposal for $D=7$. The third example dealt with a modular graph function and how to reproduce its Laurent polynomial. A last family of examples was a generalisation of the $D^6R^4$ function in $D=10$ presented in appendix~\ref{app:reg}.

While the method seems powerful and convenient for producing formal solutions to the equations, there are a number of important and interesting points that require further investigation for bringing the method to its full power. Besides the question of convergence of the Poincar\'e sum, these are
\begin{enumerate}
\item If the original automorphic differential equation comes equipped with boundary conditions such as compatibility with perturbation theory, these boundary conditions must be rephrased for the new differential equation for the seed $\sigma$. The direct translation is not obvious and a general procedure will probably rely on a solution to the second point below. In the examples in this paper we have given a heuristic set of boundary conditions based on strong couplings limits at the level of the seed.

\item The boundary conditions and the physical content of the Poincar\'e sum $\varphi (g) =\sum_{\gamma} \sigma(\gamma g)$ are expressed through the Fourier expansion of $\varphi$. Even though $\sigma$ was solved using Fourier expansion, the direct translation of this into the Fourier expansion of $\varphi$ is very hard. For the case of $SL(2,\ints)$ some intuition can be gleaned by considering the form of the Fourier expansion of the solution given in general in~\eqref{eq:PSsumFE}. These expressions, though explicit, seem impossible to evaluate for the seeds $\sigma$ that we found for the $D^6R^4$ solution\footnote{
Plugging in our solution~\eqref{eq:solZ10} and~\eqref{eq:solNZ10} this agrees with~\cite[Eq.~(B.13)]{Green:2014yxa}. In both cases, there appears to be a problem with absolute convergence as the na\"ive separate evaluation of the $m=0$ term in the zero mode $f_0(y)$ leads to $\sum_{d>0} S(0,0;d) d^{-2s} =\zeta(2s-1)/\zeta(2s)\to \infty$ for $s\to 1$ from the linear term in $y$ in $c_0(y)$. This problem is related to the lack of convergence of the Poincar\'e series for this term alone that was mentioned below~\eqref{eq:solNZ10} as this term normally produces the second constant term of the Eisenstein series $E_s(z)$.} and also in the case of modular graph functions in section~\ref{sec:MGF} we had to deal with divergent sums.

However, we can see from~\eqref{eq:PSsumFE} that the identity element in the Poincar\'e sum always yields  the Fourier mode $c_n$ of the seed and this is why we relied on the properties of this in our heuristic analysis of the boundary conditions. One obtains the same types of integral if one does the direct Fourier expansion of the solution in the $SL(5)$ case. It would be very good to develop general methods for this and also for higher rank cases.

\item Automorphic forms solving homogeneous equations can belong to principal series representations and satisfy interesting functional equations~\cite{Langlands}. Is there a similar theory underlying the solutions to the inhomogeneous equations? There is no obvious such functional relation for the one-parameter family of solutions in appendix~\ref{app:reg}. It is also not clear what the representation-theoretic meaning of the solutions is. They most probably represent vectors (or packets) in the tensor product of principal series representations.

\item The solution we found for $D^6R^4$ in $D=7$ space-time dimensions is very different from the integral formulas found in~\cite{Bossard:2015foa,Pioline:2015yea}. Since a Fourier expansion has not been achieved in those cases either, it is hard to compare the two results. The solution we constructed was based on the Laplace equation. For higher rank groups and curvature corrections one typically has more differential equations to solve and they can be of higher order in derivatives~\cite{Bossard:2015uga}, either of homogeneous or inhomogeneous type. They represent elements in the center of the universal enveloping algebra and generated the annihilator ideal for standard automorphic forms. It would be interesting to investigate these tensorial type equations for Poincar\'e sum solutions and they will constrain the homogeneous solutions further.

\end{enumerate}

\subsection*{Acknowledgements}
We would like to thank G.~Bossard, S.~Friedberg, M.B.~Green, S.D.~Miller, D.~Persson, B.~Pioline, P.~Vanhove and D.~Zagier for discussions. AK gratefully acknowledges the warm hospitality of the Max Planck Institute for Mathematics and the Hausdorff Institute in Bonn during the final stages of this work.

\appendix

\section{Particular solution of the equation for $c_N$ for $SL(5)$}
\label{app:ps}

In this appendix, we construct the particular solution for the non-zero Fourier mode $c_N(r,e_4)$ that was stated in~\eqref{eqn:nonzeromodesl5}. The equation to solve is (with $k=\gcd(N)$):
\begin{align}
\label{eq:NZ7}
    \underbrace{\left( \frac{5}{8}r^2\partial_r^2 - \frac{15}{8}r\partial_r - 4\pi^2 r^2 ||e_4^{-1}N||^2 + \Delta_{SL(4)} - \frac{42}{5} \right)}_{D}c_{N} = \underbrace{-16\pi\zeta(3)\sigma_2(k)}_{A_{k}} r^{19/5}\frac{K_1(2\pi r ||e_4^{-1}N||)}{||e_4^{-1} N||}\,.
\end{align}

\subsection{Laplacians and Bessel functions}

The $SL(4)$ invariant scalar Laplacian on $SL(4)/SO(4)$ is given by
\begin{align}
    \Delta_{SL(4)} = \frac{1}{2}g_{ik}g_{jl}\partial^{ij}\partial^{kl} - \frac{1}{8}\left( g_{ij}\partial^{ij} \right)^2 + \frac{5}{2}g_{ij}\partial^{ij}
  \label{eqn:delta4}
\end{align}
where
\begin{align}
  g = e_4 e_4^{\mathrm{T}}
  \quad{}\text{and}\quad{}
  \partial^{ij} \equiv \frac{\partial}{\partial g_{ij}}
  \quad{}\text{satifies}\quad{} 
  \partial^{ij} g_{kl} = \delta^i_k \delta^j_l + \delta^j_k \delta^i_l
\end{align}
as in~\cite[App.~A]{Obers:1999um} up to an overall normalisation consistent with our conventions.\footnote{The derivative $\partial^{ij}$ is secretly with respect to the matrix with diagonal elements rescaled by $2$ as in~\cite{Obers:1999um}, but what we need is its characteristic property when differentiating the symmetric metric $g_{kl}$.} 
We introduce some helpful notation
\begin{subequations}
\begin{align}
    &u = ||e_4^{-1} N || = \sqrt{N^T g^{-1} N} = \sqrt{N_i g^{ij} N_j} \,,\\
  &x = 2\pi r ||e_4^{-1} N || = 2\pi r u 
\end{align}
\end{subequations}
and define the functions
\begin{align}
\label{eq:fsab}
  f^{\alpha \beta}_{s} = r^\alpha u^\beta K_s\,,\quad
  f'^{\alpha \beta}_{s} = r^\alpha u^\beta K'_s\,,\quad
  f''^{\alpha \beta}_{s} = r^\alpha u^\beta K''_s\,,
\end{align}
so that the prime only affects the Bessel function. All these functions have the Bessel part evaluated at $x=2\pi r u$, e.g. $f'^{\alpha\beta}_s \equiv r^\alpha u^\beta K'_s(x)$. Note that the right-hand side of the differential equation~\eqref{eq:NZ7} is of the form $A_k\, f_1^{19/5,-1}$ and so contains a function in the class~\eqref{eq:fsab} that we shall use below to construct an ansatz for the solution.

We record some helpful identities
\begin{subequations}
\label{eq:udiff}
\begin{align}
  &g_{ij}\partial^{ij}u = -u \label{eqn:u1}\,, \\
  &g_{ik}g_{jl}\partial^{ij}\partial^{kl} u = 9u  \label{eqn:u2} \,,\\
  &g_{ik}g_{jl} \left( \partial^{ij} u \right)\left( \partial^{kl} u \right) = u^2 \label{eqn:u3} \,,
\end{align}
\end{subequations}
where repeated indices are summed over. 
We also record how the differential operators in~\eqref{eq:NZ7} act on a function $f_s^{\alpha\beta}$ from~\eqref{eq:fsab}. By using the identities \eqref{eq:udiff} one can derive the following relations
\begin{subequations}
\begin{align}
  r^2 \partial_r^2 f^{\alpha \beta}_s =& \alpha \left( \alpha - 1 \right)f^{\alpha \beta}_s + 2\alpha x f'^{\alpha \beta}_s + x^2 f''^{\alpha \beta}_s \,,\\
  r\partial_r f^{\alpha \beta}_s =& \alpha f^{\alpha \beta}_s + x f'^{\alpha \beta}_s \,,\\
  g_{ik}g_{jl}\partial^{ij}\partial^{kl} f^{\alpha \beta}_s =& \left( \beta^2 + 8\beta \right) f^{\alpha \beta}_s + \left( 2\beta + 9 \right) x f'^{\alpha \beta}_s + x^2 f''^{\alpha \beta}_s \,,\\
  \left( g_{ij}\partial^{ij} \right)^2 f^{\alpha \beta}_s =& \beta^2 f^{\alpha \beta}_s + \left( 2\beta + 1 \right) x f'^{\alpha \beta}_s + x^2 f''^{\alpha \beta}_s \,,\\
  g_{ij}\partial^{ij} f^{\alpha \beta}_s =& -\beta f^{\alpha \beta}_s - x f'^{\alpha \beta}_s \, .
\end{align}
\end{subequations}
Acting with $D$ from~\eqref{eq:NZ7} on a term $f_s^{\alpha \beta}$  then gives
\begin{equation}
\label{eq:Dfs}
        D f^{\alpha \beta}_{s} = x^2 f''^{\alpha \beta}_s +  \frac{5\alpha + 3\beta}{4}  x f'^{\alpha \beta}_s+\left(   \frac58\alpha\left( \alpha - 4 \right) + \frac{3}{8}\beta\left( \beta + 4 \right) - x^2 -\frac{42}5\right) f^{\alpha \beta}_s \,.        
\end{equation}

This expression can be further reduced to a more algebraic equation by using properties of the Bessel function $K_s(x) = K_{-s}(x)$. The first identity is the modified Bessel equation
\begin{align}
    x^2 K''_s(x) + x K'_s(x) - (x^2 + s^2) K_s(x) = 0 \quad\text{or}\quad x^2 f''^{\alpha\beta}_s + x f'^{\alpha\beta}_s - (x^2 + s^2)f^{\alpha\beta}_s = 0
\end{align}
that can be used to eliminate the second derivative of the Bessel function without changing the order $s$ of the Bessel function. Moreover, we have the recursive Bessel relation
\begin{align}
\label{eq:Besselrel}
  x K'_s(x) = s K_s (x) - x K_{s+1} (x) \quad\text{or}\quad  x f'^{\alpha\beta}_s = s f_s^{\alpha\beta} -2\pi f_{s+1}^{\alpha+1,\beta+1}
\end{align}
that can be used to replace first derivatives of the Bessel function at the cost of changing the order.

Let us first apply the modified Bessel equation to~\eqref{eq:Dfs}. This yields
\begin{align}
       D f_s^{\alpha\beta} 
        &=  \frac{5\alpha + 3\beta-4}{4} x f'^{\alpha \beta}_s  +\left(   \frac58\alpha\left( \alpha - 4 \right) + \frac{3}{8}\beta\left( \beta + 4 \right) +s^2 -\frac{42}5\right) f^{\alpha \beta}_s \,.
\end{align}
Applying then the Bessel relation~\eqref{eq:Besselrel} to this leads to
\begin{align}
\label{eq:algD}
D f_s^{\alpha\beta} = \left(   \frac58\alpha\left( \alpha - 4 \right) + \frac{3}{8}\beta\left( \beta + 4 \right) +s^2 +\frac{5\alpha + 3\beta-4}{4}s-\frac{42}5\right) f^{\alpha \beta}_s -2\pi \frac{5\alpha + 3\beta-4}{4}f_{s+1}^{\alpha+1,\beta+1}\,,
\end{align}
which implements the differential operator $D$ as a completely algebraic operation on the space of functions $\{ f_s^{\alpha\beta}\}$.

As a check on the rewriting of the differential operator, one can work out the $SL(5)$ Laplacian (i.e.\ removing the $-42/5$ from $D$) on the Fourier mode~\eqref{eq:R4Fn} of the $R^4$ correction term which corresponds to $\alpha=7/5$, $\beta=-1$ and $s=1$ to obtain the eigenvalue $-12/5$ as needed. The last term with shifted order drops out in this case as needed. We also note that the following functions are in the kernel of $D$: 
\begin{align}
\label{eq:kerD}
 D f_{s(\beta)}^{(4-3\beta)/5, \beta} =0
\end{align}
with $s(\beta) = \sqrt{(50-12\beta-3\beta^2)/5}$. This is not necessarily a complete description of the kernel.

There is one more algebraic relation that we record for special low order
\begin{align}
\label{eq:K210}
 K_2(x) = \frac{2}{x} K_1(x) + K_0(x) \quad\text{or}\quad  f^{\alpha\beta}_2 = \frac{1}{\pi} f_1^{\alpha-1,\beta-1} + f_{0}^{\alpha \beta}\,.
\end{align}

\subsection{Particular solution by recursion}

In order to find a particular solution to \eqref{eqn:nonzeromodesl5} we make the ansatz 
\begin{align}
    c_{N}^{(\mathrm{p})} = A_k \sum_i B_i r^{\alpha_i} u^{\beta_i} K_{s_i} = A_k \sum_i B_i f^{\alpha_i \beta_i}_{s_i} \quad{}\text{with}\quad{} B_i \in \reals
  \label{eqn:ansatz}
\end{align}
where the number of terms in the sum is to be determined and we take out the overall numerical factor $A_k$. From the $SL(2)$ example in~\eqref{eq:solNZ10} we expect that a small and finite number suffices. 

Plugging the ansatz \eqref{eqn:ansatz} into the differential equation \eqref{eq:NZ7}, we get upon use of~\eqref{eq:algD}
\begin{align}
&\sum_i B_i \bigg\{\left(   \frac58\alpha_i\left( \alpha_i - 4 \right) + \frac{3}{8}\beta_i\left( \beta_i + 4 \right) +s_i^2 +\frac{5\alpha_i + 3\beta_i-4}{4}s_i-\frac{42}5\right) f^{\alpha_i \beta_i}_{s_i} \nn\\
&\hspace{80mm} -2\pi \frac{5\alpha_i + 3\beta_i-4}{4}f_{s_i+1}^{\alpha_i+1,\beta_i+1}\bigg\}= f_1^{19/5,-1}\,.
\end{align}

Our strategy will be to solve this recursively such that we always generate the right-hand side from the non-order preserving term in~\eqref{eq:algD}. The right-hand side $f_1^{19/5,-1}$ can be generated by acting on $f_{0}^{14/5,-2}$ from the last term in~\eqref{eq:algD}. Since
\begin{align}
-\frac{1}{2\pi} D f_0^{14/5,-2} = \frac{12}{2\pi} f_0^{14/5,-2} +f_1^{19/5,-1} \,,
\end{align}
we start the recursion with $-\frac{1}{2\pi} f_0^{14/5,-2}$ 
to generate the original right hand side. 
This produces a new right-hand side involving $-\frac{12}{2\pi} f_0^{14/5,-2}$ that we now cancel by the same method. Since
\begin{subequations}
\begin{align}
D f_{-1}^{9/5,-3} &= -10 f_{-1}^{9/5,-3} +2\pi f_0^{14/5,-2}\,\\
\label{eq:Dfm2}
D f_{-2}^{4/5,-4} &= 6\pi f_{-1}^{9/5,-3}\,,
\end{align}
\end{subequations}
we can find a linear combination of the two functions that produces exactly the new right-hand side, \textit{viz.}
\begin{align}
D \left(-\frac{12}{(2\pi)^2} f_{-1}^{9/5,-3} -\frac{40}{(2\pi)^3} f_{-2}^{4/5,-4}\right) = -\frac{12}{2\pi} f_0^{14/5,-2} 
\end{align}
Here, it is crucial that in~\eqref{eq:Dfm2} no eigenvalue term is produced and our method terminates.

Thus, altogether we obtain the relation
\begin{align}
D \left( -\frac{1}{2\pi} f_0^{14/5,-2} -\frac{12}{(2\pi)^2} f_{-1}^{9/5,-3} -\frac{40}{(2\pi)^3} f_{-2}^{4/5,-4}\right)  = f_1^{19/5,-1}\,,
\end{align}
leading to the particular solution for the Fourier mode of the seed
\begin{align}
c_N^{\mathrm{(p)}} (r, e_4) &=  32\pi^2\zeta(3) \sigma_2(k)  r^{24/5} \left( \frac{K_0(2\pi r ||e_4^{-1}N||)}{(2\pi r ||e_4^{-1}N||)^2}+12\frac{K_1(2\pi r ||e_4^{-1}N||)}{(2\pi r ||e_4^{-1}N||)^3}+40\frac{K_2(2\pi r ||e_4^{-1}N||)}{(2\pi r ||e_4^{-1}N||)^4} \right)\nn\\
&=32\pi^2\zeta(3) \sigma_2(k)  r^{24/5} \bigg\{ \left(1+ \frac{40}{(2\pi r ||e_4^{-1}N||)^2}\right)\frac{K_0(2\pi r ||e_4^{-1}N||)}{(2\pi r ||e_4^{-1}N||)^2} \nn\\
&\hspace{40mm}+ \left(12 + \frac{80}{(2\pi r ||e_4^{-1}N||)^2}\right)\frac{K_1(2\pi r ||e_4^{-1}N||)}{(2\pi r ||e_4^{-1}N||)^3} \bigg\}
\end{align}
upon substituting back the constants and definitions, together with the symmetry $K_{-s}=K_s$. In the final rewriting, we have used the identity~\eqref{eq:K210} to eliminate the $K_2$ Bessel function and make the solution more similar to~\eqref{eq:IIBnz} that arose in the ten-dimensional type IIB case.

Similar to the solution of the homogeneous equation in the type IIB case we can expect to have at least a homogeneous solution involving $K_{7/2}$. This can be manufactured using one of the functions in~\eqref{eq:kerD} using $\beta=-\tfrac52$ or $\beta=-\tfrac32$, i.e.\ the functions
\begin{align}
\label{eq:hom7}
f_{7/2}^{23/10,-5/2} \quad\textrm{and}\quad f_{7/2}^{17/10,-3/2}
\end{align}
are homogeneous solutions to~\eqref{eq:NZ7}. They are used in the main text to propose the seed  of the $D^6R^4$ threshold function for $D=7$ space-time dimensions.

\section{Regularisation of inhomogeneous Laplace equations in $D=10$}
\label{app:reg}

In this appendix, we generalise the solution to the inhomogeneous Laplace equation~\eqref{eq:D6R410} found in section~\ref{sec:10D}. The generalisation consists in deforming the inhomogeneous equation to 
\begin{align}
\left(\Delta -(3-\epsilon)(4-\epsilon)\right) f_\epsilon(z) = -4\zeta(3)\zeta(3+2\epsilon) E_{3/2}(z) E_{3/2+\epsilon}(z)
\end{align}
The value $\epsilon=0$ corresponds to the actual $D^6R^4$ correction in $D=10$. The reason for this generalisation is that we would like to avoid the problem with the apparent singular behaviour when carrying out the Poincar\'e sum. Note that the generalisation is exact in $\epsilon$.

This equation can be treated by the same method as in section~\ref{sec:10D}. We fold the sum on the deformed Eisenstein series on the right-hand side to replace it by $y^{3/2+\epsilon}$ and obtain the following equations for the zero mode and non-zero modes of $f_\epsilon(z) = d_0(y) + \sum_{n \neq 0} d_n(y) e^{2\pi i n x}$:
\begin{subequations}
\begin{align}
\label{eq:def0}
\left(y^2 \partial_y^2 - (3-\epsilon)(4-\epsilon)\right) d_0(y) &= -4 \zeta(3)\zeta(3+2\epsilon) y^{3+\epsilon} -\frac43\pi^2\zeta(3+2\epsilon) y^{1+\epsilon},\\
\label{eq:defN0}
\left( y^2\partial_y^2 - 4\pi^2n^2y^2 -(3-\epsilon)(4-\epsilon)\right) d_n(y) &= -16\pi \zeta(3+2\epsilon) y^{2+\epsilon} |n| \sigma_{-2}(|n|) K_1(2\pi|n| y).
\end{align}
\end{subequations}

The solution to the zero mode equation~\eqref{eq:def0} is given by
\begin{align}
\label{eq:defsol0}
d_0(y) = \frac{2}{3-6\epsilon} \zeta(3) \zeta(3+2\epsilon) y^{3+\epsilon} +\frac{\pi^2}{9(1-6\epsilon)} \zeta(3+2\epsilon) y^{1+\epsilon}.
\end{align}
Here, we have fixed the solution of the homogeneous equation to zero in the same way as in section~\ref{sec:10D}.

The homogeneous form of~\eqref{eq:defN0} for the non-zero modes has the solution $y^{1/2} K_{7/2-\epsilon}(2\pi|n| y)$. Combining it with a particular solution we find
\begin{align}
d_n(y) &= \frac{8 \zeta(3+2\epsilon) \sigma_{-2}(|n|) y^{1+\epsilon}}{1-4\epsilon^2}  \bigg[ 
  \left(1-2\epsilon + \frac{10-4\epsilon}{\pi^2(|n|y)^2}\right) K_0(2\pi|n|y) &\nn\\
&\quad\quad + \left( \frac{6-\epsilon}{\pi |n| y} +\frac{10-4\epsilon}{\pi^3 |n|^3 y^3}\right)K_1(2\pi |n| y)
 -\frac{10-4\epsilon}{\Gamma\left(\tfrac72-\epsilon\right)(\pi |n| y)^{1/2+\epsilon}} K_{7/2-\epsilon}(2\pi |n| y)
\bigg],
\end{align}
where we have fixed the homogeneous solution such that the most singular terms at strong coupling ($y\to 0$) are absent. This is in correspondence with the choice for $d_0(y)$ at weak coupling ($y\to\infty$).  One can check that the above solution tends to~\eqref{eq:solNZ10} when $\epsilon\to 0$.

The advantage of this solution is that the Poincar\'e sum of the zero mode~\eqref{eq:defsol0} converges for $\epsilon>0$.

\end{document}